\documentclass[aps,pra,showpacs,twocolumn]{revtex4}

\usepackage{graphicx}
\usepackage{amsmath}
\usepackage{amssymb}
\usepackage{mathrsfs}
\usepackage{amsthm}
\usepackage{bm}
\usepackage{url}
\usepackage[T1]{fontenc}
\usepackage{csquotes}
\MakeOuterQuote{"}

\newtheoremstyle{note}
  {\topsep}               
  {\topsep}               
  {}                      
  {\parindent}            
  {\itshape}              
  {.}                     
  {5pt plus 1pt minus 1pt}
  {}

\theoremstyle{note}

\newtheorem{lemma}{Lemma}

\theoremstyle{definition}

\theoremstyle{remark}

\def\vec#1{\bm{#1}} 
\newcommand{\mrm}[1]{\mathrm{#1}}
\newcommand{\tr}{\operatorname{tr}}
\newcommand{\Tr}{\operatorname{Tr}}
\providecommand{\det}{\operatorname{det}}
\newcommand{\Det}{\operatorname{Det}}
\newcommand{\diag}{\operatorname{diag}}

\newcommand{\rmi}{\mathrm{i}}

\newcommand{\rmE}{\mathrm{E}}
\newcommand{\rmd}{\mathrm{d}}

\newcommand{\be}{\begin{equation}}
\newcommand{\ee}{\end{equation}}
\newcommand{\ba}{\begin{align}}
\newcommand{\ea}{\end{align}}

\def\<{\langle}  
\def\>{\rangle}  

\newcommand{\dket}[1]{| #1\>\!\>}

\newcommand{\dbra}[1]{\<\!\< #1|}

\def\outer#1#2{|#1\>\<#2|}       
\newcommand{\dinner}[2]{\<\!\< #1| #2\>\!\>}

\newcommand{\douter}[2]{| #1\>\!\>\<\!\< #2|}

\newcommand{\norm}[1]{\parallel\!#1\!\parallel}

\newcommand{\mse}{\mathcal{E}}
\newcommand{\mhs}{\mathcal{E}_{\mathrm{HS}}}

\newcommand{\msb}{\mathcal{E}_{\mathrm{SB}}}
\newcommand{\mtr}{\mathcal{E}_{\tr}}
\newcommand{\barcal}[1]{\bar{\mathcal{#1}}}

\newcommand{\bid}{\bar{\mathbf{I}}}

\def\eqref#1{\textup{(\ref{#1})}}  
\newcommand{\eref}[1]{Eq.~\textup{(\ref{#1})}}
\newcommand{\Eref}[1]{Equation~\textup{(\ref{#1})}}
\newcommand{\esref}[1]{Eqs.~\textup{(\ref{#1})}}

\newcommand{\fref}[1]{Fig.~\ref{#1}}
\newcommand{\Fref}[1]{Figure~\ref{#1}}

\newcommand{\Fsref}[1]{Figures~\ref{#1}}

\newcommand{\sref}[1]{Sec.~\ref{#1}}
\newcommand{\Sref}[1]{Section~\ref{#1}}
\newcommand{\ssref}[1]{Secs.~\ref{#1}}

\newcommand{\lref}[1]{Lemma~\ref{#1}}
\newcommand{\Lref}[1]{Lemma~\ref{#1}}

\newcommand{\cref}[1]{Conjecture~\ref{#1}}
\newcommand{\Cref}[1]{Conjecture~\ref{#1}}

\newcommand{\aref}[1]{Appendix~\ref{#1}}
\newcommand{\asref}[1]{Appendices~\ref{#1}}

\newcommand{\rcite}[1]{Ref.~\cite{#1}}
\newcommand{\rscite}[1]{Refs.~\cite{#1}}

\begin{document}
\title{Quantum state estimation with  informationally overcomplete measurements}
\author{Huangjun Zhu}
\email{hzhu@pitp.ca}
\affiliation{Perimeter Institute for Theoretical Physics, Waterloo, Ontario, Canada N2L 2Y5}
\affiliation{Centre for Quantum Technologies, %
National University of Singapore, Singapore 117543, Singapore}
\affiliation{NUS Graduate School for Integrative Sciences and
Engineering, Singapore 117597, Singapore}




\begin{abstract}

We study informationally overcomplete  measurements for quantum state estimation so as to clarify their tomographic significance  as compared with minimal  informationally complete measurements. We show that informationally overcomplete measurements can improve the
tomographic efficiency significantly over minimal measurements when the states of interest
have  high purities. Nevertheless, the efficiency is still too limited to be satisfactory with respect to figures of merit based  on  monotone Riemannian metrics, such as the Bures metric and quantum Chernoff metric.
In this way, we  also  pinpoint the limitation of nonadaptive measurements and motivate the study of more sophisticated measurement schemes. In the course of our study,  we introduce  the best linear unbiased estimator  and show  that it is equally efficient as  the maximum likelihood estimator  in the large-sample limit. This estimator may significantly outperform the canonical linear estimator for  states with high purities. It is expected to  play an important role in  experimental designs and adaptive quantum state tomography besides its significance to the current study.

\end{abstract}

\date{\today}
\maketitle

\section{\label{sec:introduction}Introduction}

Quantum state estimation is a procedure for inferring the state of a
quantum system from generalized measurements \cite{PariR04, LvovR09}.
A central problem in  quantum state estimation is to determine the
state of a quantum system as efficiently as possible with suitable
measurements and data processing. In practice, the set of accessible
measurements is usually determined by experimental settings, which
are not easy to modify. Given an ensemble of
identically prepared quantum systems, the simplest measurement
schemes consist of identical and independent measurements on
individual copies. A measurement is \emph{informationally complete} (IC) if every state is determined
completely by the measurement statistics \cite{Prug77, Busc91,
DAriPS04}.  Such a  measurement has at least $d^2$ outcomes for a $d$-level quantum system. An IC measurement is \emph{minimal} if it has exactly $d^2$ outcomes and  \emph{informationally overcomplete} (IOC) otherwise. A prominent example of minimal IC measurements are \emph{symmetric informationally complete} (SIC) measurements \cite{Zaun11, ReneBSC04, ScotG10, ApplFZ14G}, whereas measurements composed of complete sets of \emph{mutually unbiased bases} (MUB) \cite{Ivan81,WootF89,  DurtEBZ10} are IOC. Note, however, that the later measurements are minimal IC among combinations of projective measurements. Another example of IOC measurements is the \emph{covariant measurement}, whose outcomes consist of all pure states weighted by the Haar measure. The efficiencies of minimal IC measurements and special IOC measurements, such as mutually unbiased measurements have been studied extensively in the literature \cite{Ivan81,WootF89, JameKMW01, RehaH02, RehaEK04,EmbaN04,Scot06,RoyS07, DuSPD06, LingSLK06, BurgLDG08, LingLK08, AdamS10, DurtEBZ10, BaieP10,TeoZE10, ZhuE11, Zhu12the, PetzR12E,ApplFZ14G}. Still we often hear the basic question: which one  is more efficient for state estimation, SIC or MUB? Even little is known about general IOC measurements \cite{BurgLDG08,Tasn08}. In particular, it is not so clear what is the efficiency limit of IOC measurements, whether such measurements are useful in improving the tomographic efficiency over minimal IC measurements, and when and to what extent if the answer is positive. These general questions   are the main motivations behind the present study, which  extends some recent work presented  in the author's thesis \cite{Zhu12the}.

To answer the questions raised in the previous paragraph, we need to choose suitable figures of merit and
estimators. Among common choices of figures of merit are the mean-square error (MSE) with respect to the Hilbert-Schmidt (HS) distance and its generalization---weighted mean-square errors (WMSEs), which include the mean-square Bures distance (MSB) as a special case.  In traditional linear state tomography, the estimator is constructed in terms of measurement frequencies and reconstruction operators \cite{PariR04,Scot06, RoyS07, ZhuE11,Zhu12the}. The set of canonical
reconstruction operators is optimal if these operators are required to be independent of the measurement
statistics  \cite{Scot06,ZhuE11,Zhu12the}. However, such a choice generally cannot  make full use of the information provided by an IOC measurement.
To make a fair comparison among various measurements entails considering reconstruction operators that are optimal in the pointwise sense,
which may depend on the measurement statistics. A similar problem has
been addressed by D'Ariano and Perinotti \cite{DAriP07} (see also \rscite{BisiCDF09P,BisiCDF09I}), who derived
the set of optimal reconstruction operators with respect to the MSE in estimating certain observables.
The situation is not so clear concerning other figures of merit, such as the WMSE corresponding to a generic weighting matrix, say, the MSB. Furthermore, several basic questions are not well understood. For
example, by how much can the efficiency  be improved with
the optimal reconstruction operators instead of  the canonical
choice?

In this paper, we  determine the set of optimal reconstruction
operators in the pointwise sense and derive the \emph{best linear unbiased estimator} (BLUE), using the MSE matrix as a
benchmark. The BLUE is as efficient as the \emph{maximum likelihood estimator} (MLE)  \cite{Fish22,Fish25, Hrad97, PariR04} in the large-sample limit.
Compared with the ML approach, our approach has the merit
that it is parametrization independent and is thus often much easier
to work with and easier for deriving analytical results.  Also, it can help clarify the differences between
canonical state reconstruction and optimal  reconstruction since
the two alternatives are treated in a unified framework. Our approach
is  simpler than the one studied in
\rcite{DAriP07}, but the result has wider applicability.
In particular, it is applicable for studying tomographic efficiencies with respect to a variety of figures of merit, including various WMSEs, such as the MSE and MSB, as well as the volume of the uncertainty ellipsoid, which is pertinent to constructing good region estimators \cite{ChriR12, Blum12, ShanNSL13}.  Furthermore,
the current work provides a stepping stone  for exploring quantum state estimation with more sophisticated measurement schemes,  such as adaptive measurements \cite{Zhu12the}. What is more remarkable, certain results presented here prove to be useful for studying information theoretic analogs of uncertainty and complementarity relations \cite{Zhu14IC}.

Based on the above work, we show that covariant measurements are optimal among all
nonadaptive measurements in minimizing the average WMSE based on any
unitarily invariant distance, including the MSE and the MSB.
Compared with minimal IC  measurements, covariant measurements can
improve the tomographic efficiency significantly when the states of interest have high purities. However, the efficiency is still too limited to be satisfactory with respect to the
 scaled MSB, which  diverges at
the boundary of the state space in the large-sample limit. This
divergence is also persistent for any scaled WMSE based on a monotone
Riemannian metric \cite{Petz96,PetzS96,BengZ06book} as long as the
measurement is nonadaptive, in sharp contrast with the intuitive
belief that states with high purities are easier to estimate than
states with low purities. These general conclusions are further corroborated by extensive study of qubit state estimation with IOC measurements.
Our work not only clarifies the power of IOC measurements compared with minimal IC measurements, but also pinpoints the limitation of nonadaptive measurements, thereby motivating the  exploration of  more sophisticated measurement schemes, which we hope to address in the future.

The rest of the paper is organized as follows. In \sref{sec:OptRecIOC}, we discuss optimal state reconstruction for IOC measurements in comparison with canonical reconstruction and illustrate the matter with SIC and MUB measurements. In \sref{sec:CovMeas}, we clarify the  efficiency advantage of IOC measurements over minimal IC measurements as well as the limitation of nonadaptive measurements. In \sref{sec:QubitIOC},  we focus on qubit state estimation with IOC measurements. \Sref{sec:summary} summarizes this paper.

\section[Optimal state reconstruction]{\label{sec:OptRecIOC}Optimal state reconstruction for informationally
overcomplete measurements}

In this section we study optimal state reconstruction for general IC measurements with emphasis on IOC measurements, in preparation for the discussions in the rest of the paper.
In particular, we determine the BLUE for any  IC measurement and  show that it is as efficient as the MLE in the large-sample limit as long as the states of interest are not on the boundary of the state space. As an application of  this result, we clarify the relative merits of SIC and MUB measurements in quantum state estimation. To this end, we first need to review the basic framework of linear state tomography \cite{PariR04,Scot06, RoyS07, ZhuE11, Zhu12the}.
\subsection{\label{sec:LinearST}Linear state tomography}

A  generalized measurement is composed of a set of outcomes
represented mathematically by positive operators $\Pi_\xi $ that sum up
to the identity  1 \cite{NielC00book} (this simplified description is adequate for us since we are only concerned with the measurement statistics, not the state after the measurement). Given an unknown  state $\rho$, the
probability of obtaining the outcome $\Pi_\xi $ is given by the Born
rule: $p_\xi =\tr(\Pi_\xi \rho)$. Following the convention in \rscite{ZhuE11, Zhu12the} (see also \rscite{DAriPS00,DAriP07}), the probability can be expressed as an inner product $\dinner{\Pi_\xi }{\rho}$ between the operator kets $\dket{\Pi_\xi}$ and $\dket{\rho}$, where the double ket notation is used to distinguish them from ordinary kets.
A measurement is IC if every  state is determined  by the measurement statistics, namely, the set of probabilities $p_\xi$.
This amounts to the requirement that
the \emph{frame superoperator}
\begin{equation}\label{eq:FrameSO1}
\mathcal{F}=d\sum_\xi \frac{\douter{\Pi_\xi }{\Pi_\xi }}{\tr(\Pi_\xi)}
\end{equation}
is invertible \cite{Scot06,ZhuE11, DAriP07}, where the factor
$d$ is introduced for the convenience of later discussions.

For an IC measurement,
we can find  a set of
reconstruction operators $\Theta_\xi$
with the property
$\sum_\xi\douter{\Theta_\xi}{\Pi_\xi}=\mathbf{I}$,
where $\mathbf{I}$ is the identity superoperator. Then  any state can be recovered from the set of
probabilities $p_\xi$ as $\rho=\sum_\xi p_\xi\Theta_\xi$. In practice, the probabilities $p_\xi$ need to be replaced by the frequencies $f_\xi$  since the number $N$ of measurements is finite. The estimator based on these frequencies
$\hat{\rho}=\sum_\xi f_\xi\Theta_\xi$ is thus different from the true
state.  Nevertheless, the requirement $\sum_\xi\douter{\Theta_\xi}{\Pi_\xi}=\mathbf{I}$ on the reconstruction operators guarantees that the estimator is unbiased, that is, $\rmE(\hat{\rho})=\rho$. In general, these
frequencies obey a multinomial distribution with the scaled covariance
matrix (that is the covariance matrix multiplied by the number  of measurements) $\Sigma_{\xi\zeta}=p_\xi\delta_{\xi\zeta}-p_\xi p_\zeta$. The scaled MSE
matrix (or covariance matrix) of the estimator $\hat{\rho}$ is then determined by the formula of error propagation \cite{ZhuE11},
\begin{align}\label{eq:CovarianceMatrix}
\mathcal{C}(\rho)&=\sum_{\xi,\zeta}\dket{\Theta_\xi}\Sigma_{\xi\zeta}\dbra{\Theta_\zeta}\nonumber\\
&=\sum_\xi\dket{\Theta_\xi}p_\xi\dbra{\Theta_\xi}-\douter{\rho}{\rho}.\end{align}
Denote by $\Delta\rho=\sqrt{N}(\hat{\rho}-\rho)$\label{sym:Deltarho}
the scaled deviation of the estimator from the true state. Then the
scaled MSE with respect to the HS distance reads
\begin{align}\label{eq:MSEg}
\mse(\rho)&:=\rmE(\norm{\Delta\rho}^2_{\mathrm{HS}})
=\Tr\{\mathcal{C}(\rho)\}\nonumber\\
&\hphantom{:}=\sum_\xi
p_\xi\tr\bigl(\Theta_\xi^2\bigr)-\tr(\rho^2).
\end{align}
Here "$\Tr$"\label{sym:Tr}  denotes the trace of a superoperator,
and "$\tr$"\label{sym:tr} of an ordinary operator.

The set of reconstruction operators is not unique except for a minimal IC
measurement, such as a SIC measurement. In linear state tomography, usually the set of reconstruction operators, once chosen, is independent of the measurement
statistics.  In that case,
the set of \emph{canonical reconstruction
operators}
\begin{equation}\label{eq:CanonicalR}
\dket{\Theta_\xi}=\frac{d\mathcal{F}^{-1}\dket{\Pi_\xi}}{\tr(\Pi_\xi)}
\end{equation}
is the best choice  in the sense of
minimizing the MSE averaged over unitarily equivalent true states \cite{Scot06,RoyS07,ZhuE11,Zhu12the}. The resulting estimator is called \emph{canonical linear estimator} (CLE). The situation is different if reconstruction operators are allowed to depend on the measurement statistics, which is the focus of the next section.

\subsection{\label{sec:BLUE}Best linear unbiased estimator}
 In this section we  determine the set of optimal reconstruction operators in the
pointwise sense and derive the BLUE.

The following lemma is crucial to achieving our goal. Its proof is relegated to
\aref{app:lem:OptimalInverse}.
\begin{lemma}\label{lem:OptimalInverse}
Suppose $A$ and $B$ are two $m\times n$  matrices such that $A
B^\dag$ is the projector onto the support of $B^\dag$ (that is the range of $B$). Then $AA^\dag\geq (BB^\dag)^{+}$, and the
inequality is saturated if and only if $A=
B^{\dag+}=(BB^\dag)^{+}B$. If, in addition, $A B^\dag=1$, then
$AA^\dag\geq (BB^\dag)^{-1}$, and the inequality is saturated  if
and only if $A=(BB^\dag)^{-1}B$.
\end{lemma}
Here $A^+$ denotes the (Moore-Penrose) pseudoinverse of $A$ (the
arithmetics of pseudoinverses can be found in \rcite{Bern05book}).

Given \eref{eq:CovarianceMatrix}, \Lref{lem:OptimalInverse} applied to the matrices $\bigl(\dket{\Theta_1}p_1^{1/2}, \dket{\Theta_2}p_2^{1/2},\ldots)$
and $\bigl(\dket{\Pi_1}p_1^{-1/2}, \dket{\Pi_2}p_2^{-1/2},\ldots)$ with respect to a suitable operator basis
yields
\begin{equation}\label{eq:CovarianceMatrixOpt}
\mathcal{C}(\rho)\geq\mathcal{F}(\rho)^{-1}-\douter{\rho}{\rho},
\end{equation}
where
\begin{equation}\label{eq:FrameSOopt}
\mathcal{F}(\rho)=\sum_{\xi}\dket{\Pi_\xi}\frac{1}{p_\xi}\dbra{\Pi_\xi}
\end{equation}
is also called the frame superoperator, which generalizes the
definition in \eref{eq:FrameSO1}. To avoid unnecessary technicality, we assume that $\rho$ has full rank and thus $p_\xi>0$ for all $\xi$; rank-deficient states can be treated in suitable  limits.
The inequality is saturated if and
only if the reconstruction operators are of the form
\begin{equation}\label{eq:OptimalR}
\dket{\Theta_\xi}=p_\xi^{-1}\mathcal{F}(\rho)^{-1}\dket{\Pi_\xi},
\end{equation}
in which case we get the BLUE along with the scaled MSE matrix
\begin{equation}\label{eq:BlueMSEmatrix}
\mathcal{C}(\rho)=\mathcal{F}(\rho)^{-1}-\douter{\rho}{\rho}.
\end{equation}
According to the Aitken theorem, a generalization of the Gauss--Markov theorem, the BLUE is a special instance of \emph{weighted linear least-square estimators} for which the weighting matrix is the inverse of the covariance matrix of the measurement statistics \cite{Bos07} (note that the weighting matrix here is different from the one in the definition of the WMSE).

The scaled WMSE of the BLUE for a given weighting matrix $\mathcal{W}$ reads
\begin{equation}\label{eq:BlueWMSE}
\mse_{\mathcal{W}}(\rho)=\Tr\bigl\{\mathcal{W}\mathcal{F}(\rho)^{-1}\bigr\}-\dbra{\rho}\mathcal{W} \dket{\rho}.
\end{equation}
It reduces to the scaled MSE (with respect to the HS distance) when $\mathcal{W}$ is the identity,
\begin{equation}\label{eq:BlueMSE}
\mse(\rho)=\Tr\bigl\{\mathcal{F}(\rho)^{-1}\bigr\}-\tr(\rho^2).
\end{equation}
The volume of the scaled uncertainty ellipsoid is given by
\begin{align}\label{eq:BlueVol}
\mathcal{V}(\rho)&=V_{d^2-1}\sqrt{\bar{\Det}\{\mathcal{C}(\rho)\}}\nonumber\\
&=V_{d^2-1}\sqrt{\bar{\Det} \{\mathcal{F}(\rho)^{-1}-\douter{\rho}{\rho}\}},
\end{align}
where
\begin{equation}
V_{d^2-1}=\frac{\pi^{(d^2-1)/2}}{\Gamma(\frac{d^2+1}{2})}
\end{equation}
is the volume of the ($d^2-1$)-dimensional unit ball, and $\bar{\Det}(\mathcal{O})$ denotes the determinant of the restriction of $\mathcal{O}$ onto the space of  traceless Hermitian operators. All superoperators in this paper of which we need to evaluate $\bar{\Det}$ are supported on this space. In particular, this is the case for $\mathcal{C}(\rho)$, as we shall see shortly.

The inequality in \eref{eq:CovarianceMatrixOpt} implies that the BLUE is  optimal not only in minimizing the MSE but also in minimizing any other cost function that is monotonic increasing in the MSE matrix, such as various WMSEs and the volume of the uncertainty ellipsoid. This observation is  crucial to  investigating the efficiency advantage of the optimal state reconstruction over canonical reconstruction. It is also indispensable  for clarifying  the questions of whether and to what extent  IOC measurements are helpful in improving the tomographic efficiency over minimal IC measurements. Furthermore, the formulas for the BLUE and its associated MSE matrix can serve as a benchmark  for selecting more efficient measurement schemes, thereby providing a stepping stone for studying experimental designs and adaptive quantum state tomography \cite{Zhu12the}.

When $\rho$ is the completely mixed state, \esref{eq:FrameSOopt} and~\eqref{eq:OptimalR}
reduce to \esref{eq:FrameSO1} and~\eqref{eq:CanonicalR}, respectively, and it follows that the set of
canonical reconstruction operators and the CLE are optimal. This
observation implies that the canonical reconstruction is optimal in
minimizing the WMSE averaged over unitarily equivalent  states
as long as the weighting matrix is state independent. In the case
the weighting matrix is a constant matrix, this  conclusion reduces to the one
of Scott that the set of canonical reconstruction operators is optimal in  minimizing the average MSE \cite{Scot06} (see \sref{sec:LinearST}).

\begin{figure}
  \centering
\includegraphics[width=8cm]{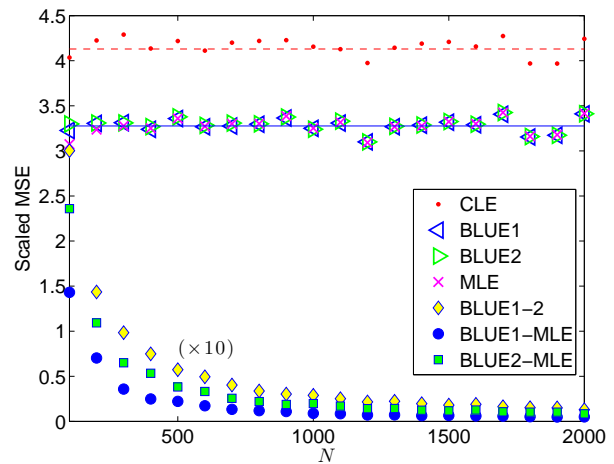}
  \caption[Tomographic efficiencies of BLUE and MLE]
  {\label{fig:QubitBlueML} (Color online) Tomographic efficiencies of the CLE, BLUE, and MLE.    The scaled MSEs of these estimators are determined by numerical simulation of  the cube measurement (see \sref{sec:QubitIOC}) on a qubit state with   random Bloch vector $\vec{s}=(0.6886, 0.1137, -0.5025)$. Each data point is an average over 1000 repetitions. BLUE1  assumes the knowledge of the true state in computing the reconstruction operators, while BLUE2  uses frequencies instead of probabilities in relevant formulas.  The theoretical scaled MSEs of the CLE and BLUE
are shown as  dashed line and solid line, respectively.
  Also plotted  are  pairwise scaled MSEs (multiplied by a factor of 10 for ease of viewing) among BLUE1, BLUE2, and MLE.   The figure indicates that the  three estimators  are almost equally efficient as long as $N$ is not too small.
  }
\end{figure}

Meticulous readers may have noticed that the optimal reconstruction
operators depend on the true state, which is usually unknown. To
remedy this problem, we may replace the true state in the relevant
formulas with an estimator obtained from  another reconstruction
scheme, canonical reconstruction for instance. Alternatively, we may just replace  probabilities $p_\xi$ with frequencies $f_\xi$ in  \esref{eq:FrameSOopt} and~\eqref{eq:OptimalR}.
In that case,  the final estimator is no longer linear in the frequencies. So strictly speaking, the BLUE is not a linear estimator in the usual sense. Nevertheless,  the resulting estimator is almost as good as the theoretical BLUE as long as $N$ is not too small.  To see this, note that for an IC measurement, any reasonable estimator, such as the CLE, will converge to the true state in the large-$N$ limit. Therefore, intuitively, the reconstruction operators based on the estimator will also converge to the theoretical optimal reconstruction operators.
Numerical calculation indicates that the MSE between the approximate BLUE and the theoretical BLUE decreases approximately as $1/N^2$, in sharp contrast with the scaling law  $1/N$ of the MSE between each estimator and the true state. For most values of $N$ of practical interest, there is almost no difference between the two estimators, as illustrated in \fref{fig:QubitBlueML} along with  the CLE and MLE  (see \sref{sec:OptML} and \aref{sec:MLrecon}).
Therefore, the BLUE is useful not only to theoretical study but also to practical applications.

For the convenience of subsequent discussions,  here we collect several basic
properties of the frame superoperator and the optimal reconstruction
operators,
\begin{subequations}\label{eq:OptimalRabc}
\begin{gather}
\mathcal{F}(\rho)\dket{\rho}=\dket{1},\quad
\mathcal{F}(\rho)^{-1}\dket{1}=\dket{\rho},  \label{eq:OptimalRa} \\
\tr(\Theta_\xi)=1, \label{eq:OptimalRb}\\
\sum_\xi \tr(\Pi_\xi)\Theta_\xi=1.  \label{eq:OptimalRc}
\end{gather}
\end{subequations}
\Eref{eq:OptimalRa} follows from the definition of
$\mathcal{F}(\rho)$; \eref{eq:OptimalRb} can be derived by
multiplying both sides of \eref{eq:OptimalR} with $\dbra{1}$ and
applying \eref{eq:OptimalRa}; \eref{eq:OptimalRc} follows from the
requirement $\sum_\xi\douter{\Theta_\xi}{\Pi_\xi}=\mathbf{I}$ and thus holds
for any set of reconstruction operators, regardless of whether it is
optimal or not.

According to \esref{eq:BlueMSEmatrix} and~\eqref{eq:OptimalRa},
$\dket{1}$ is a null eigenvector of $\mathcal{C}(\rho)$; that is,
$\mathcal{C}(\rho)$ is supported on the space of traceless Hermitian
operators as claimed before. Let $\bid$ denote the projector onto this space and define  $\barcal{F}(\rho)$ as the projection of
$\mathcal{F}(\rho)$ onto this space,
\begin{equation}\label{sym:FrameSOoptTL}
\barcal{F}(\rho):=\bid
\mathcal{F}(\rho)\bid=\sum_{\xi}\dket{\bar{\Pi}_\xi}\frac{1}{p_\xi}\dbra{\bar{\Pi}_\xi},
\end{equation}
where $\bar{\Pi}_\xi=\Pi_\xi-\tr(\Pi_\xi)/d$.
Then we can deduce from \eref{eq:OptimalRabc} that
$\mathcal{C}(\rho)\barcal{F}(\rho)=\bid$, which implies that
$\mathcal{C}(\rho)$ is the inverse of $\barcal{F}(\rho)$ in the
space of traceless Hermitian operators. Consequently,
\begin{equation}\label{eq:BlueMSE2}
\begin{aligned}
\mathcal{C}(\rho)&=\barcal{F}(\rho)^{+},& \mse_{\mathcal{W}}(\rho)&=\Tr\bigl\{\mathcal{W}\barcal{F}(\rho)^{+}\bigr\},\\
\mse(\rho)&=\Tr\bigl\{\barcal{F}(\rho)^{+}\bigr\},&
\mathcal{V}(\rho)&=V_{d^2-1}[\bar{\Det}\{\barcal{F}(\rho) \}]^{-1/2}.
\end{aligned}
\end{equation}
Comparison with \eref{eq:BlueMSEmatrix} yields
\begin{equation}\label{eq:FrameSOproj}
\barcal{F}(\rho)^{+}=\mathcal{F}(\rho)^{-1}-\douter{\rho}{\rho}.
\end{equation}
This simple formula is quite useful in later study.

In the rest of this section, we briefly discuss the problem of state
reconstruction when the measurement is not IC \cite{TeoZER11}. This problem is also
relevant to studying IOC measurements, such
as mutually unbiased measurements, since many of them are combinations of informationally incomplete measurements.

For an informationally incomplete measurement, it is generally
impossible to infer the true state accurately even if the sample
size  is arbitrarily large. Nevertheless, the projection of the true
state onto the reconstruction subspace, the space spanned by the
$\Pi_\xi$, can be determined  in the asymptotic limit. Let
$\rho_{\mathrm{R}}$ and $\mathcal{C}_{\mathrm{R}}(\rho)$ be the
restrictions  of the true state and the scaled MSE matrix onto the
reconstruction subspace. Then using a similar argument that leads to \eref{eq:CovarianceMatrixOpt}, we find
\begin{equation}
\mathcal{C}_{\mathrm{R}}(\rho)\geq\mathcal{F}(\rho)^{+}-\douter{\rho_\mathrm{R}}{\rho_\mathrm{R}}
=\barcal{F}(\rho)^+.
\end{equation}
The inequality is saturated if and only if the reconstruction
operators are given by
\begin{equation}\label{eq:OptimalReIIC}
\dket{\Theta_\xi}=p_\xi^{-1}\mathcal{F}(\rho)^{+}\dket{\Pi_\xi},
\end{equation}
when restricted to the reconstruction subspace.

To illustrate the above idea, let us consider a rank-one projective
measurement $\{\Pi_\xi\}$ for example. Noticing that the outcomes $\Pi_\xi$ are orthogonal
projectors and that $\rho_{\mathrm{R}}=\sum_\xi p_\xi\Pi_\xi$, we get
\begin{equation}\label{eq:CovarianceOptimalIIC}
\begin{aligned}
\mathcal{C}_{\mathrm{R}}(\rho)&=\sum_{\xi}\dket{\Pi_\xi}p_\xi\dbra{\Pi_\xi}-\sum_{\xi,\zeta}\dket{\Pi_\xi}p_\xi
p_\zeta\dbra{\Pi_\zeta},\\
\mathcal{E}_{\mathrm{R}}(\rho)&=\Tr\{\mathcal{C}_{\mathrm{R}}(\rho)\}=1-\sum_{\xi}p_\xi^2.
\end{aligned}
\end{equation}

\subsection{\label{sec:SICMUB}Illustration with SIC  and MUB measurements}

To illustrate the improvement of  the BLUE over the CLE and to answer a  question  raised in \sref{sec:introduction}, here we  consider state estimation with SIC measurements and complete sets of mutually unbiased measurements. Although the main results concerning SIC and MUB presented in this section were known before, they were derived under various different assumptions scattered in the literature, and a coherent account is still lacking. We hope to bridge this gap by stating the conclusion explicitly and precisely within a unified framework.

In a $d$-dimensional Hilbert space, a SIC measurement  is composed of $d^2$ subnormalized projectors onto pure states $\Pi_\xi=|\psi_\xi\rangle\langle\psi_\xi |/d$ with equal pairwise fidelity
\cite{Zaun11, ReneBSC04},
\begin{equation} \label{eq:SICinner}
|\langle\psi_\xi|\psi_\zeta\rangle|^2=\frac{d\delta_{\xi\zeta}+1}{d+1};
\end{equation}
see \rscite{ScotG10,Zhu12the,ApplFZ14G} for the latest developments.
Two bases $\{|\psi_j\rangle\}$ and
$\{|\phi_k\rangle\}$ are mutually unbiased if all the transition
probabilities $|\langle \psi_j|\phi_k\rangle|^2$ across their basis
elements are equal to $1/d$ \cite{Ivan81,WootF89,DurtEBZ10}. In a
$d$-dimensional Hilbert space, there exist at most $d+1$
MUB; such a maximal set, if it
exists, is called complete. When $d$ is a prime power,
a complete set of MUB can be constructed explicitly
\cite{Ivan81,WootF89}; see \rcite{DurtEBZ10} for a review.
Two (rank-one) projective measurements are mutually unbiased if
their measurement bases are mutually unbiased. Applications of SIC and MUB to quantum state estimation have been studied extensively in the literature
\cite{Ivan81,WootF89,  RehaH02, RehaEK04,EmbaN04,Scot06,RoyS07, DuSPD06, LingSLK06, BurgLDG08, LingLK08, AdamS10, DurtEBZ10, BaieP10,TeoZE10, ZhuE11, Zhu12the, PetzR12E,ApplFZ14G}.

For a minimal IC measurement, the optimal reconstruction is identical with the canonical reconstruction.  The scaled MSE averaged over unitarily equivalent states is bounded below by
\begin{equation}\label{eq:MinimalICbound}
\overline{\mathcal{E}(\rho)}\geq d^2+d-1-\tr(\rho^2),
\end{equation}
and the lower bound is saturated if and only if the measurement is SIC \cite{Scot06, RoyS07, ZhuE11,Zhu12the,ApplFZ14G}. For a SIC measurement, the scaled MSE is unitarily invariant, so we have
\begin{equation}\label{eq:MSEsic}
\mathcal{E}(\rho)=\overline{\mathcal{E}(\rho)}= d^2+d-1-\tr(\rho^2).
\end{equation}
The lower bound in \eref{eq:MinimalICbound} is also applicable to IOC measurements, such as mutually unbiased measurements if canonical reconstruction is applied.  The  bound is saturated if and only if the measurement is  composed of subnormalized pure states  that form a weighted 2-design \cite{ReneBSC04,Scot06, ZhuE11,Zhu12the,ApplFZ14G},  that is, $\Pi_\xi=|\psi_\xi\rangle w_\xi \langle\psi_\xi|$ with $\sum_\xi w_\xi=d$ and
\begin{equation}\label{eq:minimal2design}
\sum_\xi w_\xi (\outer{\psi_\xi}{\psi_\xi})^{\otimes 2}=\frac{2}{d+1}P_{\mrm{s}},
\end{equation}
where $P_{\mrm{s}}$ is the projector onto the bipartite symmetric subspace. Such a measurement is called \emph{tight IC} \cite{Scot06, RoyS07,ZhuE11,Zhu12the,ApplFZ14G,Zhu14GSIC}.
In that  case, the canonical reconstruction operators have a very simple form,
\begin{equation}\label{eq:TightICreconstructionO}
\Theta_\xi=|\psi_\xi\rangle (d+1) \langle\psi_\xi|-1,
\end{equation}
and the scaled MSE is also unitarily invariant  \cite{Scot06, RoyS07, ZhuE11,Zhu12the}. Since both MUB and SIC form 2-designs, it follows that they are equally efficient with respect to the MSE under canonical reconstruction.

The situation is different if the optimal reconstruction is employed. Now the scaled MSE achievable with MUB is given by \cite{EmbaN04,RoyS07,Zhu12the}
\begin{equation}\label{eq:MSEmub}
\mathcal{E}(\rho)=\overline{\mathcal{E}(\rho)}=(d+1)\bigl[d-\tr(\rho^2)\bigl].
\end{equation}
Therefore, MUB is more efficient than SIC under the optimal reconstruction, especially for  states with high purities. This example shows that the optimal reconstruction is crucial to unleashing  the full potential of  IOC measurements and to making sensible comparison among various measurement schemes. In addition, it demonstrates that IOC measurements can  indeed improve the tomographic efficiency over minimal IC measurements in quantum state estimation, as  discussed in more detail in \ssref{sec:CovMeas} and \ref{sec:QubitIOC}.

\subsection{\label{sec:OptML}Connection with the maximum-likelihood method}
To elucidate the connection between the BLUE  and  the MLE \cite{Hrad97,
PariR04}, we need to introduce a suitable parametrization for the quantum state space.
A  convenient choice is  the affine
parametrization
\begin{equation}\label{eq:AffineParametrization}
\rho(\theta)=\frac{1}{d}+\sum_{j=1}^{d^2-1}\theta_jE_j,
\end{equation}
where the $E_j$  form an orthonormal  basis in the space of
traceless Hermitian operators.  Now the Fisher information matrix takes on the
form (see  \aref{sec:Fisher})
\begin{align}
I_{jk}(\theta)&=\sum_\xi \frac{\dinner{E_j}{\Pi_\xi}\dinner{\Pi_\xi}{E_k}
}{p_\xi}=\dbra{E_j}\mathcal{F}(\rho)\dket{E_k}\nonumber\\
&=\dbra{E_j}\barcal{F}(\rho)\dket{E_k}.
\end{align}
This equation clearly indicates that  the  superoperator $\barcal{F}(\rho)$ is essentially the Fisher
information matrix in disguise and that the BLUE is as efficient as the MLE  in the
large-$N$ limit as long as the true state is not on the boundary of the state space (see \fref{fig:QubitBlueML} for an illustration). Recall that the MSE matrix of any unbiased
estimator is lower bounded by the inverse of the Fisher information
matrix and that the  bound can be saturated asymptotically with
the MLE \cite{Fish22,Fish25,Cram46,Rao45} (see
\asref{sec:Fisher} and~\ref{sec:MLrecon}). This observation implies that the BLUE is optimal not only among linear unbiased estimators but also among all unbiased estimators in the asymptotic limit.

Alternatively, we can clarify  the connection between the BLUE and the MLE by inspecting the
likelihood functional $\mathcal{L}(\rho)$ (see \aref{sec:MLrecon}) in the large-$N$ limit. According to
\eref{eq:likelihoodLog},
\begin{align}
&\frac{\partial^2\ln \mathcal{L}(\rho) }{\partial \theta_j\theta_k}=-N\sum_\xi \frac{f_\xi}{p_\xi^2}\tr(E_j\Pi_\xi) \tr (\Pi_\xi E_k)\nonumber\\
&\approx -N\sum_\xi \frac{1}{p_\xi}\tr(E_j\Pi_\xi) \tr (\Pi_\xi E_k)\nonumber\\
&=-N\dbra{E_j}\mathcal{F}(\rho)\dket{E_k}=-N\dbra{E_j}\barcal{F}(\rho)\dket{E_k}.
\end{align}
Suppose that  the likelihood functional is maximized at  $\tilde{\theta}$.
Let  $\Delta \theta=\theta-\tilde{\theta}$; then
\begin{equation}
\begin{aligned}
\frac{1}{N}\ln \mathcal{L}(\rho)&\approx c-
\frac{1}{2}\sum_{j,k}\Delta\theta_j\Delta\theta_k
\dbra{E_j}\barcal{F}(\rho)\dket{E_k},
\end{aligned}
\end{equation}
where $c$ is a constant. Again, we find that $\barcal{F}(\rho)$ plays the role of
the Fisher information matrix.

Compared with the ML method, our approach is independent of the parametrization and is thus often more convenient to work with. In particular, it allows deriving analytical results more easily, thereby elucidating the dependence of the cost function on various parameters, such as  the dimension of the Hilbert space and the purity. Also, our approach can better clarify
the differences between canonical state reconstruction and optimal
 reconstruction as well as the differences between minimal IC measurements and IOC measurements. In addition, it is quite helpful for studying
adaptive measurements and quantum precision limit \cite{Zhu12the}. The
drawback of our approach is that the optimal reconstruction
operators need to be chosen adaptively, and it is not easy to  take
into account naturally the positivity constraint on the density
operators. Depending on the situation, one alternative may be preferable to the other, and a judicious choice is crucial to  simplifying
the problem.

\section{\label{sec:CovMeas}Tomographic significance and limitation of IOC measurements}
In this section we investigate the tomographic efficiency  of IOC measurements in comparison with minimal IC
measurements, so as to answer the questions of when and to what extent IOC measurements are  advantageous over minimal IC measurements. Our study also clarifies the limitation of  nonadaptive measurements for quantum state estimation. As we shall see
shortly, covariant measurements  play a crucial role in
understanding the tomographic significance of IOC
measurements, although it is not practical to implement them in practice.
Most previous studies on covariant measurements focused on
pure-state models \cite{Haya98}. Our study fills the gap in the case of  mixed states.

\subsection{Optimality of the covariant measurement}

Suppose $\barcal{F}_1(\rho)$ and $\barcal{F}_2(\rho)$ are the
 Fisher information matrices associated with  two given IC measurements.
If the two measurements are performed with probabilities $p_1$ and
$p_2=1-p_1$, then the Fisher information matrix is a convex
combination,
\begin{equation}
\barcal{F}(\rho)=p_1\barcal{F}_1(\rho)+p_2\barcal{F}_2(\rho).
\end{equation}
Since the function $1/x$ is operator convex over the
interval $(0,\infty)$ \cite{Bhat97}, it follows that
\begin{equation}
\mathcal{C}(\rho)\leq p_1
\mathcal{C}_1(\rho)+p_2\mathcal{C}_2(\rho),\quad
\mathcal{E}(\rho)\leq p_1
\mathcal{E}_1(\rho)+p_2\mathcal{E}_2(\rho).
\end{equation}
Taking average over unitarily equivalent states yields
\begin{equation}
 \overline{\mathcal{E}(\rho)}\leq p_1
\overline{\mathcal{E}_1(\rho)}+p_2\overline{\mathcal{E}_2(\rho)}.
\end{equation}
As a consequence, $\overline{\mathcal{E}(\rho)}\leq
\overline{\mathcal{E}_1(\rho)}=\overline{\mathcal{E}_2(\rho)}$ if
the two given measurements are  unitarily equivalent. In other
words, the average MSE never increases  by combining unitarily
equivalent measurements.  Given that the set of optimal measurements contains at least one measurement that is composed of subnormalized pure states, we conclude that the average MSE is minimized by the
covariant measurement. By the same token, so is the average WMSE
based on any unitarily invariant distance, such as the Bures
distance.

In addition to minimizing average WMSEs based on various unitarily invariant distances, the covariant measurement is also optimal in minimizing the average log volume of the uncertainty ellipsoid. To see this,
\begin{align}
\ln \mathcal{V}(\rho)&=\ln V_{d^2-1}+\frac{1}{2}\ln \bar{\Det} \{\mathcal{C}(\rho)\}\nonumber\\
&=\ln V_{d^2-1}-\frac{1}{2}\ln \bar{\Det} \{\barcal{F}(\rho)\}\nonumber\\
&=\ln V_{d^2-1}-\frac{1}{2} \bar{\Tr} \ln \{\barcal{F}(\rho)\},
\end{align}
where "$\bar{\Tr}$" denotes the trace on the space of traceless Hermitian operators. Observing that the function $\ln (x)$ is operator concave \cite{Bhat97}, we deduce
\begin{equation}
\ln \mathcal{V}(\rho)\leq p_1\ln \mathcal{V}_1(\rho)+p_2\ln \mathcal{V}_2(\rho).
\end{equation}
Now our claim follows from the same reasoning as in the previous paragraph.

\subsection{Efficiency of the covariant measurement with canonical reconstruction}
As we have seen in the previous section, the covariant measurement sets the efficiency  limit to nonadaptive measurements, so it is crucial to understand its tomographic efficiency. Before investigating  its performance under the optimal reconstruction, it is instructive to consider the situation under the canonical reconstruction. The covariant measurement is a special instance of  \emph{isotropic measurements}, whose outcomes form not only (weighted) 2-designs, but also 3-designs \cite{ZhuE11,Zhu12the}. Under canonical reconstruction, isotropic measurements share the same covariant MSE matrix and are thus equally efficient  with respect to any  figure of merit that is a function of the MSE matrix, including various WMSEs and  the volume of the uncertainty ellipsoid. So the conclusions in this section also apply to any isotropic measurement.

To evaluate the  tomographic efficiency of the covariant measurement, which  is unitarily invariant, without loss of generality, we may
assume that $\rho$ is diagonal with eigenvalues $\lambda_1, \lambda_2, \ldots, \lambda_d$.
Under canonical reconstruction, the scaled MSE matrix (see \eref{eq:CovarianceMatrix}) associated with the covariant measurement is given by
\begin{align}\label{eq:MseMatrixIso}
\mathcal{C}(\rho)&=d\int \rmd \mu (\psi) (\dket{\Theta_{\psi} } \langle \psi|\rho|\psi\rangle \dbra{\Theta_\psi})-\douter{\rho}{\rho}\nonumber\\
&=\sum_{j,k}\mathcal{Q}_{jk}(\douter{E_{jj}}{E_{kk}})\nonumber\\
&\quad +\frac{d+1}{d+2} \sum_{j\neq k}(1+\lambda_j+\lambda_k) (\douter{E_{jk}}{E_{jk}}),
\end{align}
where the $\Theta_\psi=|\psi\rangle(d+1)\langle\psi|-1$  are  reconstruction operators (see \eref{eq:TightICreconstructionO}),  $\rmd \mu(\psi)$ is
the normalized Haar measure, $E_{jk}=|j\rangle \langle k|$, and
\begin{equation}
\mathcal{Q}_{jk}=\frac{(d+1)(1+2\lambda_j)\delta_{jk}-1-\lambda_j-\lambda_k-(d+2)\lambda_j\lambda_k}{d+2}.
\end{equation}
Define
\begin{equation}\label{eq:Ejkpm}
\begin{aligned}
E_{jk}^{+}&:=\frac{1}{\sqrt{2}}(|j\rangle\langle k|+|k\rangle\langle j|),\\
E_{jk}^{-}&:=-\frac{\rmi}{\sqrt{2}}(|j\rangle\langle
k|-|k\rangle\langle j|).
\end{aligned}
\end{equation}
Then $E_{jk}^{\pm}$ with $j\neq k$ are  eigenvectors of $\mathcal{C}(\rho)$ with eigenvalues $(d+1)(1+\lambda_j+\lambda_k)/(d+2)$.

The scaled MSE agrees with \eref{eq:MSEsic} as expected for a tight IC measurement.
The scaled MSB  reads
\begin{align}\label{eq:MSBcovCan}
\msb(\rho)&=\frac{2d^3+2d^2-3d-2}{4(d+2)}\nonumber\\
&\quad+\frac{1}{4(d+2)}\biggl[\sum_j\frac{d}{\lambda_j}+\sum_{j\neq k}\frac{2(d+1)}{\lambda_j+\lambda_k}\biggr].
\end{align}
where we have applied the formula for  the  Bures metric derived by H\"ubner \cite{Hubn92},
\begin{equation}\label{eq:BuresMetric}
D_{\mathrm{B}}^2(\rho,\rho+\rmd
\rho)=\frac{1}{2}\sum_{j,k}\frac{|\langle j|\rmd
\rho|k\rangle|^2}{\lambda_j+\lambda_k}.
\end{equation}
Note that the scaled MSB diverges at the boundary of the state space. The same is true for the scaled WMSE based on any monotone Riemannian metric because the Bures metric is minimal among such metrics \cite{Petz96,PetzS96,BengZ06book}. To see this explicitly, observe that up to a multiplicative constant a generic monotone Riemannian metric has the form
\begin{equation}\label{eq:MonotoneMetric}
D_{c}^2(\rho,\rho+\rmd
\rho)=\sum_{j}\frac{|\langle j|\rmd
\rho|j \rangle|^2}{4\lambda_j}+\sum_{j\neq k}\frac{c(\lambda_j,\lambda_k)}{4}|\langle j|\rmd
\rho|k\rangle|^2.
\end{equation}
where $c(x,y)$ is a Morozova-Chentsov function \cite{Petz96,PetzS96,BengZ06book}. The corresponding scaled WMSE is given by
\begin{align}
\mathcal{E}_c(\rho)&=\frac{2d^2-d-2}{4(d+2)}+\frac{d}{4(d+2)}\sum_j\frac{1}{\lambda_j} \nonumber\\
&\quad +\frac{d+1}{4(d+2)}\sum_{j\neq k}(1+\lambda_j+\lambda_k)c(\lambda_j,\lambda_k).
\end{align}
This equation reduces to \eref{eq:MSBcovCan} if $c(x,y)=2/(x+y)$, which corresponds to the Bures metric. For the quantum Chernoff metric \cite{AudeCMB07}, we have $c(x,y)=4/(\sqrt{x}+\sqrt{y})^2$ and
\begin{align}
\mathcal{E}_c(\rho)&=\frac{2d^2-d-2}{4(d+2)}+\frac{d}{4(d+2)}\sum_j\frac{1}{\lambda_j} \nonumber\\
&\quad +\frac{d+1}{d+2}\sum_{j\neq k}\frac{1+\lambda_j+\lambda_k}{(\sqrt{\lambda_j} +\sqrt{\lambda_k})^2}.
\end{align}

\subsection{Efficiency of the covariant measurement with optimal reconstruction}
Now let us turn to the optimal state reconstruction based on the covariant measurement. According to \eref{eq:FrameSOopt}, the frame superoperator is given by
\begin{equation}
\mathcal{F}(\rho)=d\int\rmd
\mu(\psi)\frac{1}{\langle\psi|\rho|\psi\rangle}
(\dket{\Pi_\psi}\dbra{\Pi_\psi}),
\end{equation}
where $\Pi_\psi=|\psi\rangle\langle\psi|$.  In general, it is not easy to derive an explicit formula for $\mathcal{F}(\rho)$. To understand its state dependence, it is instructive to  consider those states that are convex combinations of the
completely mixed state and a  projector state of rank $r$,
\begin{equation}\label{eq:rhors}
\rho_r(s)=\frac{s}{r}\sum_{j=1}^r |j\rangle\langle
j|+(1-s)\frac{1}{d},  \quad 1\leq r\leq d-1,\; 0\leq s\leq 1.
\end{equation}
Note, however, that we do not assume this knowledge in state reconstruction.
In this case, $\mathcal{F}\bm{(}\rho_r(s)\bm{)}$ has the
form
\begin{align}\label{eq:Frs}
\mathcal{F}\bm{(}\rho_r(s)\bm{)}&=a \mathcal{P}_1+b\mathcal{P}_2+c\mathcal{P}_3
+\sum_{j,k}\mathcal{M}_{jk}\douter{E_{jj}}{E_{kk}},
\end{align}
where $\mathcal{P}_1,\mathcal{P}_2,\mathcal{P}_3$ are projectors,
\begin{align}
\mathcal{P}_1&=\sum_{j\neq
k=1}^{r}\douter{E_{jk}}{E_{jk}},\quad \mathcal{P}_3=\sum_{j\neq k=r+1}^{d}\douter{E_{jk}}{E_{jk}}, \nonumber \\
 \mathcal{P}_2&=\sum_{j=1}^r \sum_{k=r+1}^d \bigl(\douter{E_{jk}}{E_{jk}}+\douter{E_{kj}}{E_{kj}}\bigr),
\end{align}
and
\begin{equation}
\mathcal{M}_{jk}=\begin{cases}
                        (1+\delta_{jk})a & \mbox{if}\quad 1\leq j,k\leq r,\\
                        (1+\delta_{jk})c & \mbox{if} \quad r+1\leq j,k\leq d,\\
                        b                & \mbox{otherwise}.
                      \end{cases}
\end{equation}
The three parameters $a$, $ b$, and  $c$ are determined by the formulas $a=g_{20}$, $b=g_{11}$, and $c=g_{02}$, where
\begin{align}
g_{jk}&=\frac{2dr\Gamma(d+1)}{\Gamma(r+j)\Gamma(d-r+k)}\nonumber\\
&\quad\times \int_0^{\pi/2}\rmd
\alpha\frac{(\cos\alpha)^{2r-1+2j}(\sin\alpha)^{2d-2r-1+2k}}{ds(\cos\alpha)^2+r(1-s)},
\end{align}
which can be evaluated by applying the formula
\begin{align}
&\int_0^{\pi/2}\rmd
\alpha\frac{\cos\alpha(\sin\alpha)^{2m+1}}{(\cos\alpha)^2+u}\nonumber\\
&=\frac{1}{2}(1+u)^{m}\ln\frac{1+u}{u}-\frac{1}{2}\sum_{n=0}^{m-1}\frac{(1+u)^n}{m-n},\quad
u>0
\end{align}
after replacing $(\cos\alpha)^2$ with $1-(\sin\alpha)^2$. The Fisher
information matrix $\barcal{F}\bm{(}\rho_r(s)\bm{)}$ has the same
form as $\mathcal{F}\bm{(}\rho_r(s)\bm{)}$, except that
$\mathcal{M}$ is replaced by $\barcal{M}:=\bid\mathcal{M}\bid$.

Calculation shows that $\barcal{M}$ has $r-1$ eigenvalues equal to
$a$, $d-r-1$ eigenvalues equal to $c$, and one eigenvalue equal to
\begin{equation}\label{eq:eigLast}
\beta=\frac{(r+1)(d-r)a+r(d-r+1)c-2r(d-r)b}{d}.
\end{equation}
Note that $E_{jk}^{\pm}$ for $j\neq k$ are eigenvectors of  $\mathcal{F}$
and  $\barcal{F}$, and that the common eigenvalue is one of the
three choices $a, b, c$ depending on the values of $j$ and $k$.
We deduce that  $\barcal{F}$ has four distinct eigenvalues $a, b, c$, and
$\beta$ with multiplicities $r^2-1$, $2r(d-r)$, $(d-r)^2-1$, and 1,
respectively (the eigenvalue corresponding to the null eigenvector $\dket{1}$ is excluded here).

According to \eref{eq:BlueMSE2}, the scaled MSE is given by
\begin{equation}\label{eq:MSEcov}
\mse\bm{(}\rho_r(s)\bm{)}=\frac{r^2-1}{a}+\frac{2r(d-r)}{b}+\frac{(d-r)^2-1}{c}+\frac{1}{\beta}.
\end{equation}
The scaled MSB can be determined by virtue of \eref{eq:BuresMetric} with the result
\begin{align}\label{eq:MSBcov}
\msb\bm{(}\rho_r(s)\bm{)}&=\frac{1}{4}\Bigl(\frac{r^2-1}{a\lambda_1}+\frac{4r(d-r)}{b(\lambda_1+\lambda_2)}\nonumber\\
&\quad +\frac{(d-r)^2-1}{c\lambda_2}
+\frac{d-r}{d\beta\lambda_1}+\frac{r}{d\beta\lambda_2}\Bigr),
\end{align}
where $\lambda_1=(s/r)+(1-s)/d$ and $\lambda_2=(1-s)/d$ are the two
distinct eigenvalues of $\rho$. The scaled WMSEs with respect to other monotone Riemannian metrics can  be derived in a similar manner.
The volume (with respect to the HS metric) of the scaled uncertainty ellipsoid is given by
\begin{equation}\label{eq:UellipsoidVol}
\mathcal{V}\bm{(}\rho_r(s)\bm{)}=V_{d^2-1}\bigl(a^{r^2-1}b^{2r(d-r)}c^{(d-r)^2-1}\beta\bigr)^{-1/2},
\end{equation}
along with its logarithm
\begin{align}
\ln \mathcal{V}\bm{(}\rho_r(s)\bm{)}&=\ln V_{d^2-1}-\frac{1}{2} \bigl\{(r^2-1)\ln a + [2r(d-r)]\ln b\nonumber \\
&\quad +[(d-r)^2-1]\ln c
+\ln \beta\bigr\}.
\end{align}

\begin{figure}
  \centering
\includegraphics[width=7cm]{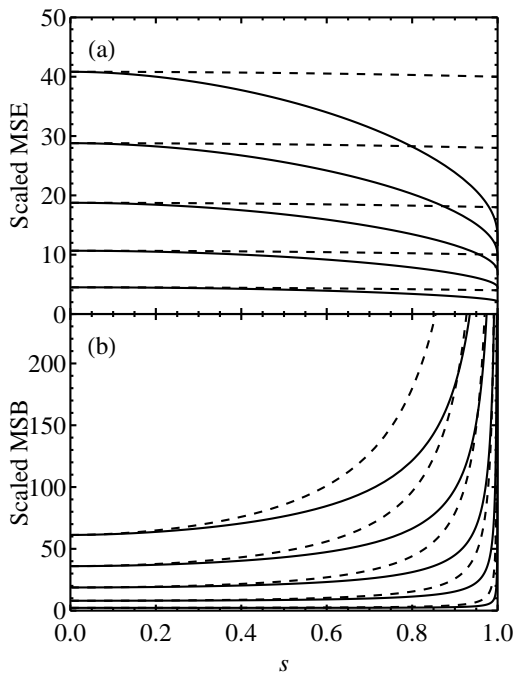}
  \caption[Tomographic efficiencies of
   covariant measurements with respect to the scaled MSE and the scaled MSB]
  {\label{fig:ocmMseMsb}Tomographic efficiencies of
   covariant measurements. The  true states have the form in
  \eref{eq:rhors} with $r=1$ and $d=2,3,\ldots,6$ (from bottom to top). The scaled MSB diverges in the limit $s\rightarrow 1$, in which case the states are rank deficient.
  For comparison,  the dashed lines show the performances of   covariant measurements under canonical linear reconstruction. In plot (a), they also represent the performances of the optimal minimal IC  measurements (that is SIC measurements) with respect to the scaled MSE.
 }
\end{figure}

\Fref{fig:ocmMseMsb} illustrates
the scaled MSE and MSB in the case $r=1$ and $d=2,3,4,5,6$. Compared
with canonical linear state tomography or minimal state tomography, optimal
state estimation with covariant measurements can improve the
efficiency significantly when the  states of interest have high purities.
Nevertheless, the efficiency is still too limited to be satisfactory
when the scaled MSB is chosen as the figure of merit.

As  $s$ approaches 1,  the state $\rho_r(s)$ turns into a  subnormalized
projector of rank $r$. When $r\geq2$, the three parameters $a,b,c$
have well-defined limits $a=r/(r+1)$, $b=1$, $c=r/(r-1)$, and so does
the scaled MSE,
\begin{equation}
\mathcal{E}\bm{(}\rho_r(1)\bm{)}=d^2+2d-1-\frac{d^2}{r}-\frac{1}{r}.
\end{equation}
When $r=1$,  the parameters $a$ and $b$  still have well-defined
limits, whereas $c$ diverges as $\ln[d/(1-s)]$. The  formula for the scaled
MSE is still applicable, except that  the derivative of
$\mathcal{E}\bm{(}\rho_r(s)\bm{)}$ with respect to $s$ can diverge.
In the pure-state limit, the scaled MSE $2(d-1)$ achieved by the covariant
measurement is equal to the corresponding value for the pure-state
model \cite{Haya98}. Compared with the scaled MSE
$d^2+d-2$ \cite{Scot06, ZhuE11,Zhu12the}  that is achievable with minimal  state tomography, it is smaller by $(d+2)/2$ times.
 Furthermore, it is minimal not only in the Bayesian sense but
also in the pointwise sense by saturating a quantum analog of
the Cram\'er-Rao bound; see \rcite{Mats02} as well as Secs. 5.3.3 and 6.2.2 of \rcite{Zhu12the}.

In the pure-state limit, the scaled MSE matrix can be determined based on
\esref{eq:BlueMSEmatrix} and~\eqref{eq:Frs}, with the result
\begin{equation}
\mathcal{C}(|1\rangle\langle1|)=\sum_{j=2}^d
\bigl(\douter{E_{1j}^+}{E_{1j}^+}+\douter{E_{1j}^-}{E_{1j}^-}\bigr).
\end{equation}
It  is a rank-$2(d-1)$ projector, in contrast with the scaled MSE matrix associated with canonical reconstruction, which has full rank in the space of traceless Hermitian operators (see \eref{eq:MseMatrixIso}).
The scaled
deviation $\Delta\rho$ has the form
\begin{equation}
\Delta\rho=\sum_{j=2}^d \bigl(x_j E_{1j}^++y_jE_{1j}^-\bigr),
\end{equation}
where $x_j, y_j$ obey a $2(d-1)$-dimensional standard isotropic
Gaussian distribution.  Since  $\Delta\rho$ has only two nonzero
eigenvalues $\pm\sqrt{\sum_{j=2}^d \bigl(x_j^2+y_j^2\bigr)/2}$, its
trace norm is proportional to the HS norm,
$\norm{\Delta\rho}_{\tr}=\norm{\Delta\rho}_{\mathrm{HS}}/\sqrt{2}$.
The scaled mean errors (not MSE) with respect to the trace distance and the  HS distance are given by
\begin{equation}
\mtr(\rho)
=\frac{1}{\sqrt{2}}\mhs(\rho)=\frac{\Gamma\bigl(d-\frac{1}{2}\bigr)}{\Gamma(d-1)}\approx\sqrt{d-1}.
\end{equation}
Compared with the result achievable with  minimal
tomography \cite{ZhuE11, Zhu12the}, the scaled mean trace distance is approximately smaller by a
factor of $4d/3\pi$ when $d\gg2$. Therefore, the efficiency advantage  of IOC measurements is more substantial  with respect to the mean trace distance in comparison with
the MSE. The contrast is even more dramatic with respect to the volume of the scaled uncertainty ellipsoid: the average  volume vanishes in the pure-state limit for the covariant measurement  but remains finite for any minimal IC measurement or any set of mutually unbiased measurements.

In sharp contrast,  the scaled MSB diverges in the limit
$s\rightarrow1$. Consequently, with respect to the Bures metric, the volume of the scaled uncertainty ellipsoid also diverges. This seemingly surprising phenomenon can be
explained as follows: the entries of $\barcal{F}$ are either finite
or logarithmically divergent  in this limit, while the entries of
the weighting matrix diverge much faster according to
\eref{eq:BuresMetric}. Recalling that the covariant measurement
minimizes the average scaled MSB among all nonadaptive measurements,
we conclude that the average scaled MSB diverges at the boundary of
the state space for all nonadaptive measurements. From the Bayesian
perspective, our analysis implies that the MSB generally decreases more slowly than the scaling law $1/N$  expected from common
statistical consideration once the prior  weight near pure states is non-negligible. For single qubit, this
phenomenon was noticed in \rcite{BagaBGM06S}.
The same conclusion also holds for any WMSE based on a monotone
Riemannian metric since the Bures
metric is  minimal among all such metrics
\cite{Petz96,PetzS96,BengZ06book}.
These observations reveal a severe
limitation of nonadaptive measurements for quantum  state estimation and the importance of exploring more sophisticated strategies, which deserve further study  \cite{Zhu12the}.

\section{\label{sec:QubitIOC}Qubit state estimation with informationally overcomplete measurements}
In this section we exemplify our general approach on IOC measurements with qubit state estimation. Our main goal is to elucidate with this simple example the efficiency limit of IOC measurements and the extent to which they are advantageous  over minimal IC measurements with respect to various figures of merit, such as  the MSE, MSB,  and the volume of the uncertainty ellipsoid. To be concrete, our discussions focus on  the covariant
measurement and measurements constructed out of platonic solids inscribed on the Bloch sphere. Nevertheless, our approach applies equally well to other measurements.
There are already many studies
on this subject \cite{RehaEK04, LingSLK06, BurgLDG08, LingLK08}, but most
theoretical works are based on numerical simulations. We have
derived several analytical results on canonical  linear state tomography in
 \rcite{ZhuE11}. Here we turn to the optimal state reconstruction in comparison with the canonical reconstruction.

\subsection{Canonical reconstruction}
Following the convention in   \rscite{ZhuE11,Zhu12the}, to each platonic
solid inscribed on the Bloch sphere, we can construct  a generalized
measurement whose outcomes correspond to the vertices of the platonic solid.
Given a platonic solid with $n$ vertices represented by $n$ unit vectors~$\vec{v}_k$, the outcomes of the corresponding measurement are given by $\Pi_k=(1+\vec{v}_k\cdot\vec{\sigma})/n$. Suppose the qubit state~$\rho$ is parametrized by the Bloch vector
$\vec{s}=(x,y,z)$; then reconstructing the  state $\rho$ is
equivalent to  reconstructing its Bloch vector $\vec{s}$.

Under canonical reconstruction, the reconstruction operators take on  the form
$\Theta_k=(1+3\vec{v}_k\cdot\vec{\sigma})/2$ according to \eref{eq:TightICreconstructionO} since the measurement corresponding to any platonic solid
is tight IC. The
scaled MSE matrix of the estimator $\hat{s}$ of the Bloch vector has the form \cite{ZhuE11,Zhu12the}
 \begin{equation}\label{eq:QubitCovarianceM}
C(\vec{s})=3-\vec{s}\vec{s}
+\frac{9}{n}\sum_{k=1}^n
\left(\vec{v}_k\cdot\vec{s}\right)
\vec{v}_k\vec{v}_k,
\end{equation}
where $\vec{s}\vec{s}$ is the dyadic composed of the vector $\vec{s}$ and itself.
 For any measurement constructed from a platonic solid other than the regular tetrahedron, the last term in the equation vanishes due to symmetry, which yields
\begin{equation}\label{eq:QubitMseMatrixIso}
C^{\mathrm{Iso}}(\vec{s})=3-\vec{s}\vec{s}.
\end{equation}
More generally, all isotropic measurements \cite{ZhuE11,Zhu12the} share the same scaled MSE matrix and are equally efficient under canonical reconstruction.
The scaled
MSE with respect to the HS distance is equal to
\begin{equation}
\mse(\rho)=\frac{1}{2}\tr\{C(\vec{s})\}=\frac{9-s^2}{2}.
\end{equation}
Here the factor $1/2$ accounts for the difference between the HS distance and the distance on the Bloch ball.
The scaled MSE is independent of the orientation of the Bloch vector, regardless of the platonic solid under consideration, as expected for any rank-one tight IC measurement.

The weighting matrix corresponding to the Bures metric is one fourth of the quantum Fisher information matrix and takes on the form
\begin{equation}
W(\vec{s})=\frac{1}{4}+\frac{\vec{s}\vec{s}}{4(1-s^2)}.
\end{equation}
The scaled MSB is thus given by
\begin{align}
\msb(\rho)&=\tr\{W(\vec{s})C(\vec{s})\}\nonumber\\
&=\frac{9}{4}+\frac{s^2}{2(1-s^2)}+\frac{9}{4n(1-s^2)}\sum_k(\vec{s}\cdot\vec{v}_k)^3.
\end{align}
Except for the SIC (tetrahedron) measurement, the last term vanishes, and we have
\begin{align}
\msb(\rho)=\frac{9}{4}+\frac{s^2}{2(1-s^2)}.
\end{align}
To derive an explicit formula for the SIC measurement, we assume that  the cube (also the octahedron) takes on the standard
orientation and that the tetrahedron is composed of four vertices of  the cube including $(1,1,1)/\sqrt{3}$. In that case,
\begin{align}\label{eq:MsbSIC}
\msb^{\mathrm{SIC}}(\rho)=\frac{9}{4}+\frac{s^2+3\sqrt{3}xyz}{2(1-s^2)}.
\end{align}
Unlike the scaled MSE, which is unitarily invariant,  the scaled MSB for given $s$ is maximized when the Bloch vector of the true state is parallel to one leg of the outcomes and minimized in the opposite situation.
The last term in the above equation vanishes after taking average over unitarily equivalent states. Therefore, all measurements constructed from platonic solids are equally efficient with respect to the average scaled MSB under canonical reconstruction. This conclusion is not as obvious as the corresponding statement concerning the scaled MSE.

The volume (with respect to the HS metric) of the scaled uncertainty ellipsoid is given by
\begin{equation}
\mathcal{V}(\rho)=\frac{4\pi}{3}\sqrt{\frac{\det\{C(\vec{s})\}}{8}};
\end{equation}
note that $V_3=4\pi/3$. Here the factor $1/8$ accounts for the difference between the HS distance and the distance on the Bloch ball as before. For isotropic measurements, it reduces to
\begin{equation}\label{eq:VolumeIso}
\mathcal{V}^{\mathrm{Iso}}(\rho)=\pi\sqrt{2(3-s^2)}.
\end{equation}
For the SIC measurement, we have
\begin{align}\label{eq:VolumeSIC}
\mathcal{V}^{\mathrm{SIC}}(\rho)&=\sqrt{\frac{2}{3}}\pi\bigl[2(x^4+y^4+z^4)+8\sqrt{3}xyz\nonumber\\
&\quad-s^4-6s^2+9\bigr]^{1/2}.
\end{align}

\subsection{Optimal reconstruction}
\begin{figure}  \centering
 \includegraphics[width=7cm]{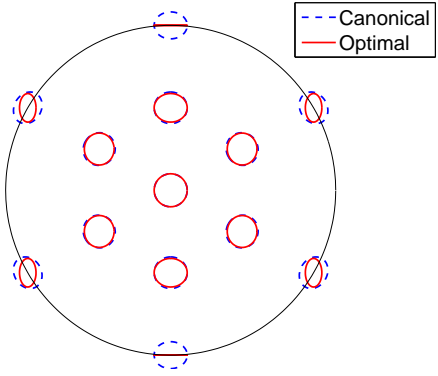}\\
  \caption[Uncertainty ellipses of
  the MUB measurement with the canonical and the optimal reconstructions]{\label{fig:UellipsesMub}
   (Color online) Uncertainty ellipses of the canonical reconstruction and the optimal reconstruction.
 The uncertainty ellipses are associated with the marginal
  distributions  on the $x$-$z$ plane of the Bloch ball resulting from mutually unbiased
  measurements on  a family of states, each repeated 300 times.   The optimal reconstruction reduces the sizes of the
uncertainty ellipses at the prize of losing the covariance property.
}
\end{figure}

Now let us turn to the optimal reconstruction.
In terms of the Bloch vector, the Fisher information matrix takes on the form
\begin{equation}
I(\vec{s})=\frac{1}{n}\sum_k \frac{1}{1+\vec{v}_k\cdot \vec{s}}\vec{v}_k\vec{v}_k.
\end{equation}
The scaled MSE matrix $C(\vec{s})$ is the inverse of $I(\vec{s})$. For the SIC measurement, it is still given by \eref{eq:QubitCovarianceM}. For the MUB measurement, we have
\begin{equation}
C(\vec{s})=3\,\diag (1-x^2,1-y^2,1-z^2).
\end{equation}
It is smaller than the scaled MSE matrix $3-\vec{s}\vec{s}$ under the
canonical reconstruction (cf.\ \eref{eq:QubitMseMatrixIso}), but is no
longer invariant under unitary transformations of the measurement
outcomes. The differences between the two reconstruction methods are
clearly reflected in the uncertainty ellipses, as illustrated in
\fref{fig:UellipsesMub}. The situations are quite similar for measurements
constructed from other platonic solids except for the tetrahedron, although the expressions of $C(\vec{s})$ can be much more complicated.

 The scaled MSEs of the measurements constructed from
the tetrahedron, octahedron, and cube are respectively given by
\begin{equation}
\begin{aligned}
\mathcal{E}^{\mathrm{SIC}}(\rho)&=\frac{9-s^2}{2},\\
\mathcal{E}^{\mathrm{MUB}}(\rho)&=\frac{3(3-s^2)}{2},\\
\mathcal{E}^{\mathrm{Cube}}(\rho)&=\frac{27-18s^2+s^4+2(x^4+y^4+z^4)}{2(3-s^2)}.
\end{aligned}
\end{equation}
 The scaled MSE
is unitarily invariant for the SIC (tetrahedron) measurement and the
MUB (octahedron) measurement, as mentioned in \sref{sec:SICMUB}.
This is not the case for the cube measurement, although it is a
combination of two tetrahedron measurements and is seemingly more
symmetric than a single tetrahedron measurement.   For given
$s$, the minimal scaled MSE $(9-s^2)(9-5s^2)/6(3-s^2)$ is attained
when $\vec{s}$ is parallel to one of the diagonals of the cube, and
the maximum $3(3-s^2)/2$ is attained when $\vec{s}$ is parallel to
one of the axes.  The average is
\begin{equation}
\overline{\mathcal{E}^{\mathrm{Cube}}(\rho)}=\frac{135-90s^2+11s^4}{10(3-s^2)}.
\end{equation}
The formulas for the MSEs of the dodecahedron measurement and
icosahedron measurement are too complicated to convey a clear
meaning; suffice it to mention that the MSEs are not unitarily
invariant in both cases, as in the case of the cube measurement. This observation reveals an intriguing feature that seems to be unique to SIC and MUB measurements, which deserves further study~\cite{Zhu14GSIC}.

The scaled MSB for the SIC measurement is still given by \eref{eq:MsbSIC}. For the MUB and cube measurements, we have
\begin{equation}
\begin{aligned}
&\msb^{\mathrm{MUB}}(\rho)=\frac{3(3-s^2)}{4}+\frac{3(s^2-x^4-y^4-z^4)}{4(1-s^2)},\\
&\msb^{\mathrm{Cube}}(\rho)=\frac{27-27s^2-2s^4}{12(1-s^2)}\\
& +\frac{6(x^4+y^4+z^4)-2(x^6+y^6+z^6)-21x^2y^2z^2}{3(3-s^2)(1-s^2)}.
\end{aligned}
\end{equation}
Taking average over unitarily equivalent states yields
\begin{equation}
\begin{aligned}
\overline{\msb^{\mathrm{SIC}}(\rho)}&=\frac{9}{4}+\frac{s^2}{2(1-s^2)},\\
\overline{\msb^{\mathrm{MUB}}(\rho)}&=\frac{9}{4}+\frac{3s^4}{10(1-s^2)},\\
\overline{\msb^{\mathrm{Cube}}(\rho)}&=\frac{945-1260s^2+413s^4-26s^6}{140(3-s^2)(1-s^2)}.
\end{aligned}
\end{equation}

The volume  of the scaled uncertainty ellipsoid of the SIC measurement is still determined  by \eref{eq:VolumeSIC}. For
MUB and cube measurements, they  are respectively given by
\begin{equation}
\begin{aligned}
\mathcal{V}^{\mathrm{MUB}}(\rho)&=\pi\sqrt{6(1-x^2)(1-y^2)(1-z^2)},\\
\mathcal{V}^{\mathrm{Cube}}(\rho)&=\frac{\pi}{3}\sqrt{\frac{2[3-(x+y-z)^2][3-(x-y+z)^2]}{3-s^2}} \\
&\times \sqrt{[3-(-x+y+z)^2][3-(x+y+z)^2]}.
\end{aligned}
\end{equation}
They are all equal to $\sqrt{6}\pi$ when $s=0$.
The averages of the log volumes over unitarily equivalent states read
\begin{equation}\label{eq:VmubCube}
\begin{aligned}
\overline{\ln \mathcal{V}^{\mathrm{MUB}}(\rho)}&=\ln(\sqrt{6}\pi)-3+\frac{3}{2}\Bigl[\ln (1-s^2)+\frac{1}{s}\ln \frac{1+s}{1-s}\Bigr],\\
\overline{\ln \mathcal{V}^{\mathrm{Cube}}(\rho)}&=\ln (3\sqrt{2}\pi)-4+\ln \frac{(1-s^2)^2}{\sqrt{3-s^2}} +\frac{2}{s}\ln \frac{1+s}{1-s}.
\end{aligned}
\end{equation}
For the SIC measurement, this average can be determined by numerical integration.

For the covariant measurement, the parameters $b$ in \eref{eq:Frs} and
$\beta$ in \eref{eq:eigLast} are now given by
\begin{equation}
b=\frac{2 s-(1-s^2)
\ln\bigl(\frac{1+s}{1-s}\bigr)}{2 s^3},\quad
\beta=\frac{-2 s+\ln\bigl(\frac{1+s}{1-s}\bigr)}{s^3}.
\end{equation}
Note that the parameters $a$ and $c$ are irrelevant here.
The Fisher information matrix takes on
the form
\begin{equation}
\barcal{F}(\rho)=b\bid+\frac{1}{2}(\beta-b)\douter{\tilde{\vec{s}}\cdot\sigma}{\tilde{\vec{s}}\cdot\sigma}.
\end{equation}
where $\tilde{\vec{s}}=\vec{s}/s$ is the normalized Bloch vector (the ambiguity at $s=0$ does not matter since $b=\beta$ in that case).
In terms of the Bloch vector, it simplifies to
\begin{equation}
I(\vec{s})=\frac{1}{2}[b+(\beta-b)\tilde{\vec{s}}\tilde{\vec{s}}].
\end{equation}
 Its inverse is the scaled  MSE matrix associated with the optimal reconstruction,
\begin{equation}
C(\vec{s})=2\Bigl[\frac{1}{b} +\Bigl(\frac{1}{\beta}-\frac{1}{b}\Bigr)\tilde{\vec{s}}\tilde{\vec{s}}\Bigr].
\end{equation}
The scaled MSE,   MSB, and the volume of the scaled uncertainty ellipsoid (with respect to the HS metric)  follow from
\esref{eq:MSEcov}, \eqref{eq:MSBcov},  and~\eqref{eq:UellipsoidVol}, respectively,
\begin{equation}\label{eq:CovMseMsbV}
\begin{aligned}
\mse(\rho)&=\frac{2}{b}+\frac{1}{\beta},\\
\msb(\rho)&= \frac{1}{b}+\frac{1}{2\beta(1-s^2)},\\
\mathcal{V}(\rho)&=\frac{4\pi}{3b\sqrt{\beta}}.
\end{aligned}
\end{equation}
Similarly, the WMSE with respect to the monotone Riemannian metric characterized by the Morozova-Chentsov function $c(x,y)$ is given by
\begin{equation}\label{eq:CovWmse}
\mathcal{E}_c(\rho)=\frac{c(\lambda_+,\lambda_-)}{2b}+\frac{1}{2\beta(1-s^2)},
\end{equation}
where $\lambda_{\pm}=(1\pm s)/2$ are the eigenvalues of $\rho$. For the Chernoff metric  $c(x,y)=4/(\sqrt{x}+\sqrt{y})^2$, it reduces to
\begin{equation}
\mathcal{E}_c(\rho)= \frac{2}{b(1+\sqrt{1-s^2})}+\frac{1}{2\beta(1-s^2)}.
\end{equation}

As comparison, in canonical linear tomography with the covariant measurement, the scaled MSE matrix $C(\vec{s})$ is equal to $3-\vec{s}\vec{s}$ as in \eref{eq:QubitMseMatrixIso} since the covariant measurement is an isotropic measurement.
Accordingly, we have
\begin{equation}
\begin{aligned}
\mse(\rho)&=\frac{9-s^2}{2}, \\
\msb(\rho)&= \frac{3}{2}+\frac{3-s^2}{4(1-s^2)},\\
 \mathcal{V}(\rho)&= \pi\sqrt{2(3-s^2)},\\
 \mathcal{E}_c(\rho)&=\frac{3}{4}c(\lambda_+,\lambda_-)+\frac{3-s^2}{4(1-s^2)}.
\end{aligned}
\end{equation}

\begin{figure}[tb]  \centering
 \includegraphics[width=7cm]{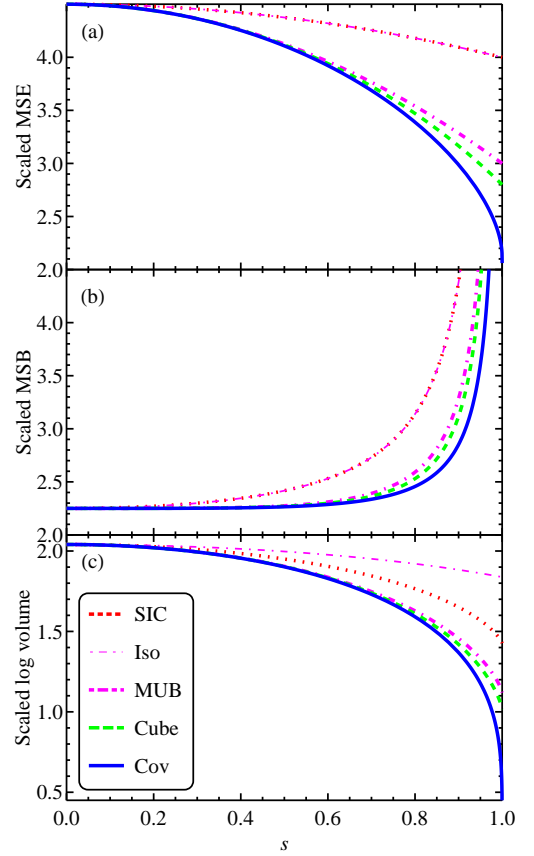}\\
  \caption[Tomographic efficiencies of  the SIC, MUB, cube, and covariant
  measurements  in qubit state estimation]{\label{fig:QubitOCM} (Color online) Tomographic efficiencies of   the SIC, MUB, cube, and covariant
  measurements  in qubit state estimation. (a) Average scaled MSE; (b) average scaled MSB; (c) average log volume (with respect to the HS metric) of the scaled uncertainty ellipsoid. The average scaled MSBs diverge in the pure-state limit for all four measurements.
For the covariant measurement, the log volume diverges to $-\infty$ (that is, the volume vanishes) in the pure-state limit.
As comparison, the curve "Iso" shows the common performance of isotropic measurements (including MUB, cube, and covariant measurements) under canonical linear reconstruction. It coincides with the curve "SIC" in plots (a) and (b)
since under canonical linear reconstruction isotropic measurements are as efficient as the SIC measurement in qubit state estimation with respect to the average scaled MSE and MSB.  }
\end{figure}

\Fref{fig:QubitOCM} shows the tomographic performances of the SIC, MUB,
cube, and covariant
  measurements in qubit state estimation with respect to the average scaled MSE, MSB, and  log volume of the scaled uncertainty ellipsoid. For all three figures of merit, the tomographic efficiencies of the four measurement schemes are monotonic increasing with the number of outcomes, the more so the higher the purities of the states of interest.  By contrast, in canonical linear state tomography,  MUB,
cube, and covariant
  measurements are as efficient as the  SIC measurement with respect to the average scaled MSE and  MSB, and  even less efficient with respect to the average log volume.
   Comparison with the scaled MSE achieved by the optimal adaptive strategy \cite{Haya97,GillM00,EmbaN04,HayaM08,Zhu12the} shows that under the optimal reconstruction
  the covariant measurement  is almost optimal in the pointwise
  sense. However,  it should be noted that this is generally not the
  case with respect to other figures of merit, such as the  scaled MSB. Also,  the situation can be very different beyond the two-level system (see Chap. 5 in \rcite{Zhu12the}). Actually, the scaled MSB diverges in the pure-state limit for the covariant measurement, although it is the most efficient among all nonadaptive measurements. The same is true for any WMSE based on a monotone Riemannian metric, as explained in \sref{sec:CovMeas}.

\section{\label{sec:summary}Summary}

We  have studied quantum state estimation with IOC measurements, motivated by  the questions of whether and to what extent IOC measurements can improve the tomographic efficiency over minimal IC measurements.
To answer these questions and to make fair comparison among various measurement schemes, we derived  the best linear unbiased estimator  and showed  that it is  as efficient as  the maximum likelihood estimator in the large-sample limit. This estimator may significantly outperform the canonical linear estimator when the states of interest have high purities. This finding is useful  not only for studying  IOC measurements but also for exploring experimental designs and adaptive quantum state estimation.

Based on the above framework, we showed that the covariant measurement is optimal
among all nonadaptive measurements in minimizing the average WMSE based on
any unitarily invariant distance, including the MSE and the MSB, as well as the average log volume of the uncertainty ellipsoid.
When the states of interest
have  high purities, IOC measurements can improve the
tomographic efficiency significantly and even change the scaling of the cost function  with the dimension of the Hilbert space. Nevertheless, the efficiency is still too limited to be satisfactory with respect to the  MSB or  the WMSE based on any other monotone Riemannian metric as
long as the measurement is nonadaptive. On the one hand, our study  clarifies the tomographic significance of IOC measurements compared with minimal IC measurements. On the other hand, it pinpoints the limitation of nonadaptive measurements and motivates the study of more sophisticated estimation
strategies based on adaptive measurements and collective
measurements~\cite{Zhu12the}, which deserve further study. In this paper, we only consider ideal measurements. It would  be desirable in the future to extend the current work to incorporate imperfection, such as detector inefficiency.

\section*{Acknowledgements}
The author is  grateful to Berthold-Georg Englert, Masahito Hayashi, and Yong Siah Teo for stimulating discussions and comments on early versions of the manuscript and to Hai Wang for comments on the proof of \lref{lem:OptimalInverse}. The author is also grateful to the referee for comments and suggestions that improve the clarity of the paper. This work is supported in part by Perimeter Institute for Theoretical Physics. Research at Perimeter Institute is supported by the Government of Canada through Industry Canada and by the Province of Ontario through the Ministry of Research and Innovation.
In the early stage, it was supported by NUS Graduate
School (NGS) for Integrative Sciences and Engineering and  Centre
for Quantum Technologies, which is a Research Centre of Excellence
funded by the Ministry of Education and National Research Foundation
of Singapore. \Fsref{fig:ocmMseMsb} and \ref{fig:QubitOCM} have been created using the LevelScheme scientific figure
preparation system \cite{Capr05}.

\bigskip
\appendix

\section{\label{app:lem:OptimalInverse}Proof of
\lref{lem:OptimalInverse}}

The idea of the proof follows from the proof of Lemma~5.1 in Chap.
VI of  \rcite{Hole82book}. Let $\bm{u}$ and $\bm{v}$ be two $m\times 1$
vectors such that $\bm{v}$ belongs to the support of $B^\dag$. Let
$\bm{a}=A^\dag\bm{u}$ and $\bm{b}=B^\dag\bm{v}$; then we have
\begin{equation}
\begin{aligned}
\bm{a}^\dag \bm{a}&=\bm{u}^\dag AA^\dag\bm{u}, \quad \bm{b}^\dag
\bm{b}=\bm{v}^\dag BB^\dag\bm{v}, \\
 \bm{a}^\dag
\bm{b}&=\bm{u}^\dag A B^\dag\bm{v}=\bm{u}^\dag \bm{v}.
\end{aligned}
\end{equation}
The Cauchy inequality applied to the equation yields
\begin{equation}
(\bm{u}^\dag AA^\dag\bm{u})(\bm{v}^\dag BB^\dag\bm{v})\geq
(\bm{u}^\dag \bm{v})^2.
\end{equation}
Setting $\bm{v}=(BB^\dag)^{+}\bm{u}$ gives rise to
\begin{equation}
\bm{u}^\dag AA^\dag\bm{u}\geq \bm{u}^\dag (BB^\dag)^{+}\bm{u},
\end{equation}
which  implies that $AA^\dag\geq (BB^\dag)^{+}$. Necessary
conditions for saturating the inequality are $A^\dag\bm{u}\propto
B^\dag(BB^\dag)^{+}\bm{u}$ and $|A^\dag\bm{u}|=
|B^\dag(BB^\dag)^{+}\bm{u}|$ for arbitrary $\bm{u}$; that is,
$A^\dag\propto B^\dag(BB^\dag)^{+}$ and  $A\propto (BB^\dag)^{+}B$.
Since $A B^\dag$ is a projector by assumption,  it follows that $A=
(BB^\dag)^{+}B$, which happens to be the pseudoinverse of $B^\dag $
\cite{Bern05book}. Now the inequality is indeed saturated.

If $A B^\dag=1$, then $(BB^\dag)$ is invertible. The second part of
the lemma follows from the fact that $(BB^\dag)^{+}=(BB^\dag)^{-1}$.

\section{\label{sec:Fisher}Fisher information and Cram\'er-Rao bound}
Fisher information \cite{Fish25} and the Cram\'er-Rao bound~\cite{Cram46, Rao45} are two
basic ingredients in statistical inference: the former quantifies
the amount of information yielded by an observation or a measurement
concerning certain parameters of interest, and  the latter
quantifies the minimal error in estimating  these
parameters.

Consider a family of probability distributions $p(\xi|\theta)$
parametrized  by  $\theta$. Our task is to estimate the value of
$\theta$ as accurately as possible based on the measurement
outcomes. Given an outcome $\xi$, the function $p(\xi|\theta)$ of
$\theta$ is called the \emph{likelihood function}. The \emph{score} is
defined as the partial derivative of the log-likelihood function
with respect to $\theta$ and reflects the sensitivity of the
log-likelihood function with respect to the variation of $\theta$.
Its first moment is zero, and the  second moment is known as the
\emph{Fisher information}
\cite{Fish25,LehmC98book},
\begin{align}\label{sym:FisherInf}
I(\theta)&=\mathrm{Var}\Bigl(\frac{\partial \ln
p(\xi|\theta)}{\partial \theta}\Bigr)= \sum_\xi
p(\xi|\theta)\Bigl(\frac{\partial \ln p(\xi|\theta)}{\partial
\theta}\Bigr)^2\nonumber\\
&=\sum_\xi\frac{1}{p(\xi|\theta)}\Bigl(\frac{\partial
p(\xi|\theta)}{\partial \theta}\Bigr)^2.
\end{align}
The Fisher information represents the average sensitivity of the
log-likelihood function with respect to the variation of $\theta$.
Intuitively, the larger the Fisher information, the better we can
estimate the value of the parameter $\theta$.

An estimator $\hat{\theta}(\xi)$ of the parameter $\theta$ is
\emph{unbiased} if its expectation value is equal to the true
parameter; that is,
\begin{equation}
\sum_\xi p(\xi|\theta) [\hat{\theta}(\xi)-\theta]=0.
\end{equation}
Taking the derivative with respect to $\theta$ and applying the
Cauchy--Schwarz inequality (using the fact that $\sum_\xi
p(\xi|\theta)=1$) yield the well-known \emph{Cram\'er-Rao
bound}
\begin{equation}\label{sym:Var}
C(\theta)=\mathrm{Var}(\hat{\theta})\geq \frac{1}{I(\theta)},
\end{equation}
which states that the MSE or variance of any unbiased estimator is bounded from
below by the inverse of the Fisher information \cite{Cram46, Rao45}.

In the multiparameter setting, the Fisher information and MSE take on  matrix forms,
\begin{equation}
\begin{aligned}
I_{jk}(\theta)&=\rmE\biggl[\biggl(\frac{\partial \ln
p(\xi|\theta)}{\partial \theta_j}\biggr)\biggl(\frac{\partial \ln
p(\xi|\theta)}{\partial \theta_k}\biggr)\biggr],\\
C_{jk}(\theta)&=\rmE[
(\hat{\theta}_j-\theta_j)(\hat{\theta}_k-\theta_k)].
\end{aligned}
\end{equation}
Accordingly, the Cram\'er-Rao bound for any unbiased estimator turns out to be a matrix
inequality,
\begin{equation}
C(\theta)\geq I^{-1}(\theta).
\end{equation}

Since the likelihood function is multiplicative, the Fisher
information matrix is additive; that is, the total Fisher
information matrix of independent measurements is equal to
the sum of the respective Fisher information matrices of individual
measurements. In particular, the Fisher information matrix of $N$
identical and independent measurements is $N$ times that of one
measurement. Accordingly, the MSE matrix of any unbiased estimator
based on $N$ measurements satisfies the inequality $C^{(N)}(\theta)\geq
1/NI(\theta)$. Thanks to Fisher's theorem \cite{Fish22,Fish25}, the
lower bound can be saturated asymptotically with the MLE under very general assumptions \cite{Bos07}. In the
large-sample scenario, the \emph{scaled MSE matrix} $NC^{(N)}(\theta)$ is generally independent of the sample size.
It is also denoted by $C(\theta)$ when there is no confusion.

In quantum state estimation, we are interested in the parameters
that characterize the state $\rho(\theta)$ of a quantum system.  To
estimate the values of these parameters, we may perform generalized
measurements. Given a measurement $\Pi$ with outcomes $\Pi_\xi$, the
probability of obtaining the outcome $\xi$ is
$p(\xi|\theta)=\tr\{\rho(\theta)\Pi_\xi\}$. The corresponding Fisher
information matrix $I_{jk}(\Pi,\theta)$ is given by
\begin{equation}
I_{jk}(\Pi,\theta)=\sum_\xi
\frac{1}{p(\xi|\theta)}\tr\biggl\{\frac{\partial
\rho(\theta)}{\partial \theta_j}\Pi_\xi\biggr\}
\tr\biggl\{\frac{\partial \rho(\theta)}{\partial
\theta_k}\Pi_\xi\biggr\}.\quad
\end{equation}
Once a measurement is chosen,  the inverse Fisher information matrix
sets a lower bound for the MSE matrix of any unbiased estimator,
which can be saturated asymptotically by the MLE, as in the
case of classical parameter estimation. It should be noted that the
bound depends on the specific measurement.

In practice, it is often more convenient to use a single number
rather than a matrix to quantify the error. A common choice is the
scaled MSE  $\tr\{C(\theta)\}$; a more general alternative is the
scaled WMSE
$\tr\{W(\theta)C(\theta)\}$, where
$W(\theta)$\label{sym:WeightMatrix} is a positive semidefinite
weighting matrix, which may depend on $\theta$. The Cram\'er-Rao bound implies
that $\tr\{W(\theta)C(\theta)\}\geq \tr\{W(\theta)
I^{-1}(\theta)\}$; again, this bound can be saturated asymptotically
with the MLE. A drawback with the MSE is that it depends on
the parametrization, which is somehow arbitrary. With a suitable
choice of the weighting matrix, the WMSE is free from this problem. For
example, the WMSEs with respect to the HS distance and Bures distance are
parametrization independent. Except when stated otherwise, the MSE concerned in the main text is defined with respect to the HS distance.

\section{\label{sec:MLrecon}Maximum-likelihood estimation}

In  ML estimation, instead of searching for a state that matches the observed
frequencies, we seek a state that maximizes the likelihood
function (or functional). The principle of ML was proposed by  Fisher \cite{Fish22} in the
1920s and has  become a basic ingredient in statistical inference.
During the past decade, it has found extensive applications in
quantum state estimation
\cite{Hrad97,PariR04,RehaHJ01,RehaHKL07,LvovR09}. In addition, it is
useful for entanglement detection \cite{BlumYE10} and
characterization \cite{ChenZW11}.

In quantum state estimation, the \emph{likelihood
functional}~\cite{Hrad97,PariR04} is defined as
\begin{equation}\label{eq:likelihood}
\mathcal{L}(\rho)=\prod_\xi p_\xi^{n_\xi},
\end{equation}
where $p_\xi=\tr(\rho\Pi_\xi)$ and $n_\xi$ are the probability and the number of times of  obtaining the
outcome $\xi$ given $N$ measurements on the  state $\rho$. In practice, it is often
more convenient to work with the \emph{log-likelihood
functional}
\begin{equation}\label{eq:likelihoodLog}
\ln\mathcal{L}(\rho)=\sum_\xi n_\xi\ln p_\xi=N \sum_\xi f_\xi\ln
p_\xi.
\end{equation}
The ML method consists in choosing a state $\hat{\rho}_{\mathrm{ML}}$ that
maximizes the  likelihood functional or, equivalently, the
log-likelihood functional, as an estimator of the true state
\cite{PariR04,LvovR09,Hrad97,RehaHJ01,RehaHKL07}. If there exists a state
that  matches the observed frequencies, then the state is also an MLE. This conclusion
is an immediate consequence of the inequality
\begin{equation} \sum_\xi f_\xi\ln p_\xi\leq \sum_\xi
f_\xi\ln f_\xi.
\end{equation}

In general, it is not easy to find a closed formula for the MLE. Fortunately, the estimator can be computed efficiently
with an algorithm proposed by Hradil \cite{Hrad97}.

\bibliographystyle{apsrev}
\bibliography{all_references}

\end{document}